\documentclass{JINST}

\title{A performance study of an electron-tracking Compton camera with a compact system for environmental gamma-ray observation}

\author{T.~Mizumoto$^a$\thanks{Corresponding
author.}, D.~Tomono$^a$, A.~Takada$^a$, T.~Tanimori$^a$, S.~Komura$^a$, H.~Kubo$^a$, Y.~Matsuoka$^a$, Y.~Mizumura$^{a,b}$, K.~Nakamura$^a$, S.~Nakamura$^a$, M.~Oda$^a$, J.~D.~Parker$^a$, T.~Sawano$^a$, N.~Bando$^c$ and A.~Nabetani$^d$\\
\llap{$^a$}Division of Physics and Astronomy, Graduate School of Science, Kyoto University, Kitashirakawa-Oiwakecho, Sakyo-ku, Kyoto, 606-8502, Japan\\
\llap{$^b$}Unit of Synergetic Studies for Space, Kyoto University, Kitashirakawa-Oiwakecho, Sakyo-ku, Kyoto, 606-8502, Japan\\
\llap{$^c$}HORIBA, Ltd, 2, Miyanohigashi, Kisshoin, Minami-ku, Kyoto, 601-8510, Japan\\
\llap{$^d$}Canon Inc., 30-2, Shomomaruko 3-chome, Ohta-ku, Tokyo, 146-8501, Japan\\
  E-mail: \email{mizumoto@cr.scphys.kyoto-u.ac.jp}}

\abstract{An electron-tracking Compton camera (ETCC) is a detector that can determine the arrival direction and energy of incident sub-MeV/MeV gamma-ray events on an event-by-event basis. It is a hybrid detector consisting of a gaseous time projection chamber (TPC), that is the Compton-scattering target and the tracker of recoil electrons, and a position-sensitive scintillation camera that absorbs of the scattered gamma rays, to measure gamma rays in the environment from contaminated soil. To measure of environmental gamma rays from soil contaminated with radioactive cesium (Cs), we developed a portable battery-powered ETCC system with a compact readout circuit and data-acquisition system for the SMILE-II experiment \cite{ueno_jinst}, \cite{mizumoto-ieee}. We checked the gamma-ray imaging ability and ETCC performance in the laboratory by using several gamma-ray point sources. The performance test indicates that the field of view (FoV) of the detector is about 1$\;$sr and that the detection efficiency and angular resolution for 662$\;$keV gamma rays from the center of the FoV is $(9.31 \pm 0.95) \times 10^{^-5}$ and  $5.9^{\circ} \pm 0.6^{\circ}$, respectively. Furthermore, the ETCC can detect 0.15$\;\mu\rm{Sv/h}$ from a $^{137}$Cs gamma-ray source with a significance of 5$\sigma$ in 13 min in the laboratory. In this paper, we report the specifications of the ETCC and the results of the performance tests. Furthermore, we discuss its potential use for environmental gamma-ray measurements.}

\keywords{Gaseous imaging and tracking detectors, Compton imaging, Dosimetry concepts and apparatus, Time projection chambers}

\begin{document}

\section{Introduction}
\label{sect:introduction}
Quantitative and quick estimation of the necessity for decontamination of radioactive cesium (Cs) contaminated soils is required in Japan. Electron-tracking Compton cameras (ETCC), which was developed mainly for sub-MeV/MeV gamma-ray astronomy \cite{takada_SMILE-I}, \cite{mizumura_SMILE-II}, is suited for monitoring gamma rays from these soils for the estimation by virtue of its imaging ability, background-rejection power, and large field of view (FoV). 

An ETCC consists essentially of two sub-detectors: a gaseous time projection chamber (TPC) and a scintillation camera (cf. top-left panel of figure \ref{photo_schematic:ETCC}). A TPC tracks Compton recoil electrons in three dimensions and detects their recoil direction, their energy, and the Compton-scattering point. On the other hand, a scintillation camera works by absorbing the scattered gamma rays and detects their energy and absorption point. The arrangement of the sub-detectors, that the low-Z gas Compton-scatterer is in front and the high-Z Photo-absorber is behind, is suitable for detection of gamma rays only from the front. Using this information, we can reconstruct the energy and arrival direction of the incident photon on an arc whose shape depends on the accuracy of the determination of the scattering angle $\phi$ (angular resolution measure (ARM)) and scattering plane (scatter plane deviation (SPD)), as shown in the top-left panel of figure \ref{photo_schematic:ETCC}. This approach contrasts the circle used by conventional Compton cameras, which can not track the Compton recoil electron. The tracking feature leads to the powerful imaging ability of ETCCs (cf. figure 2 in reference \cite{mizumura_SMILE-II}). Moreover, an ETCC has powerful background-rejection power because of the track information of the Compton-scattered electron. Furthermore, we can develop an ETCC with a large FoV. For instance, the SMILE-II flight-model (FM) ETCC has a FoV of approximately $2\pi\;\rm{sr}$ \cite{matsuoka-PSD10}. These features are suitable for environmental gamma-ray measurements.

One of our goals is to enhance the ETCC to make it a commercial product. In addition, we want to develop an ETCC that can determine a ambient dose rate in several hours with an accuracy of 10$\%$ in an environment of the ambient dose rate of 0.23$\;\mu$Sv/h, which is used as a criterion for decontamination of soils in Japan. To satisfy these requirements, we need a large-size ETCC with high detection efficiency. If the ETCC can separate gamma-ray images into 0.1$\;$sr regions and detect gamma-ray rates with $30\%$ accuracy from each region within several hours, it would be useful for detecting hot spots. As a prototype, we developed an ETCC that should work for such purposes. The first step of the development consisted of constructing the ETCC system, which uses the amplifier-shaper-discriminator (ASD) integrated circuit and ASD board for the TPC readout \cite{asd1}, \cite{asd2}, and the position-encoding system used in reference \cite{komura-ieee}. We reported the results of the initial performance tests of the ETCC in reference \cite{tomono-ieee}. In the next step, we developed a more compact, battery-powered ETCC system with a new compact readout circuit and data-acquisition system, which was developed for the SMILE-II FM ETCC \cite{mizumoto-ieee}. In the section below, we present our new ETCC system, which is dedicated for environmental gamma-ray measurements, and the result of detailed laboratory-based performance tests.

\section{Instrument}
\label{sect:instrument}
\begin{figure}[t]
 \begin{center}  
  \begin{tabular}{c}
   \begin{minipage}{0.47\hsize}
    \begin{center}
     \includegraphics[width=5.cm]{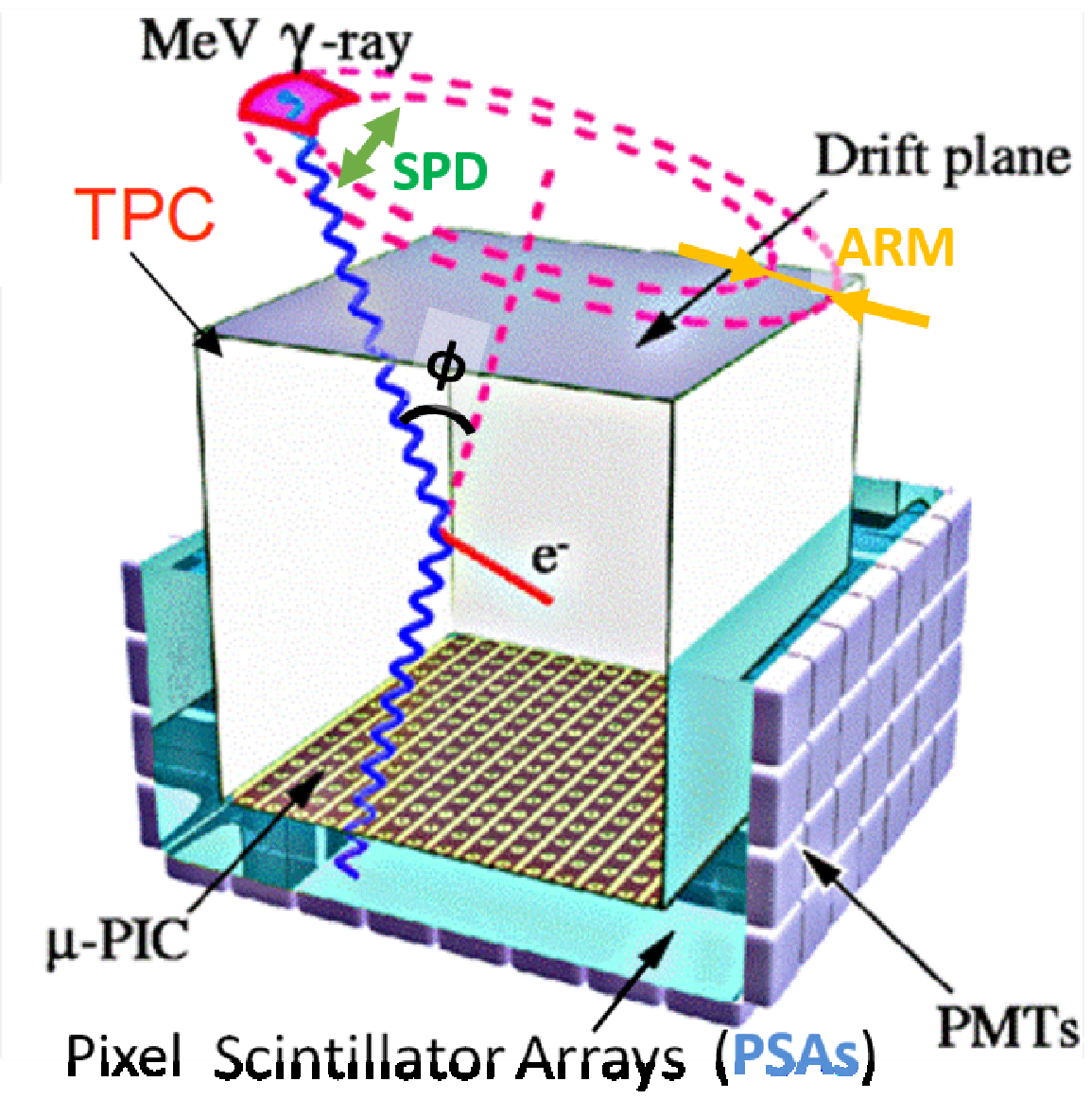}
    \end{center}
   \end{minipage}
   \begin{minipage}{0.53\hsize}
    \begin{center}
    \includegraphics[width=6.2cm]{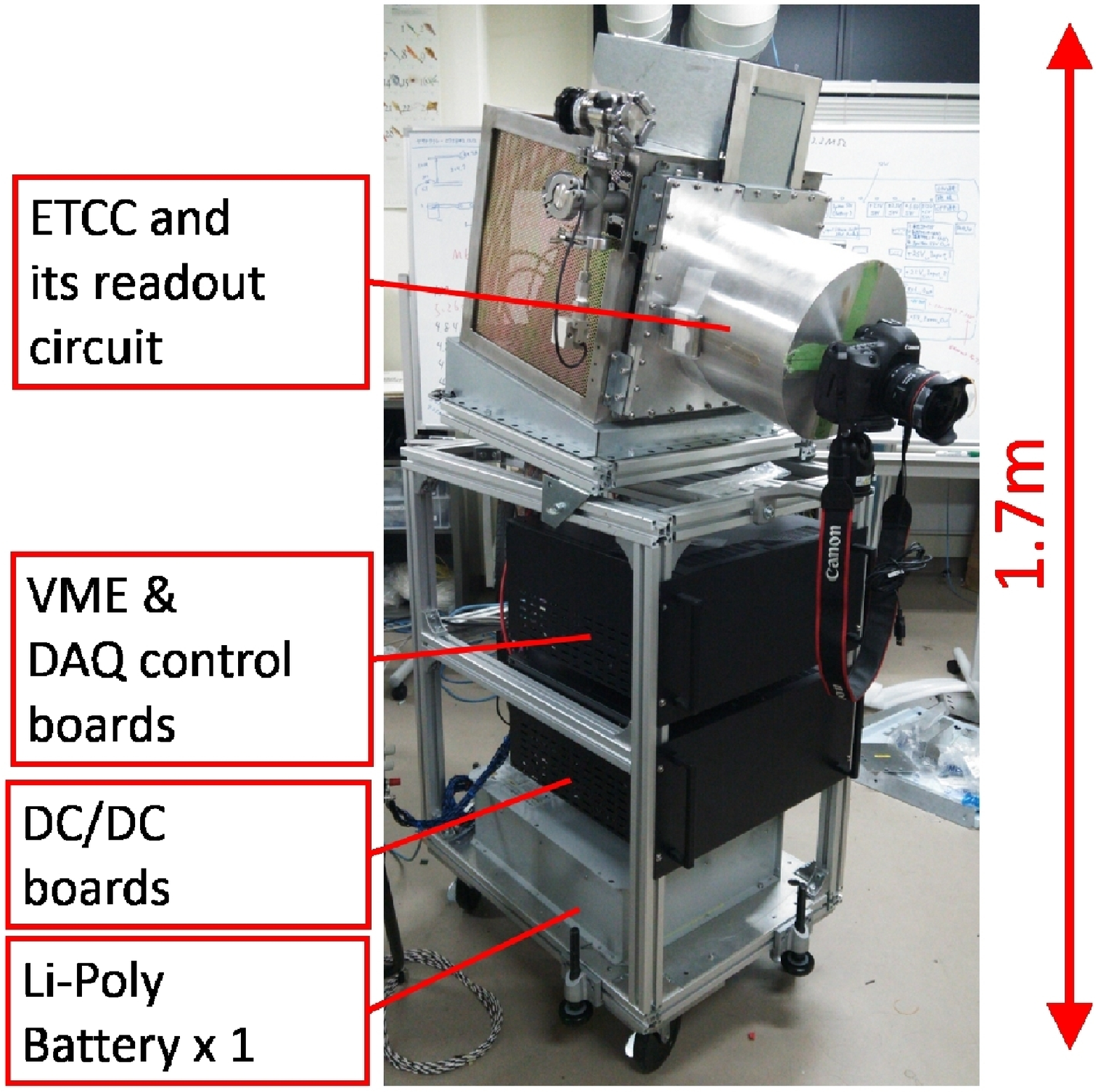}
    \end{center}
   \end{minipage}
  \end{tabular}
  \includegraphics[width=13.5cm]{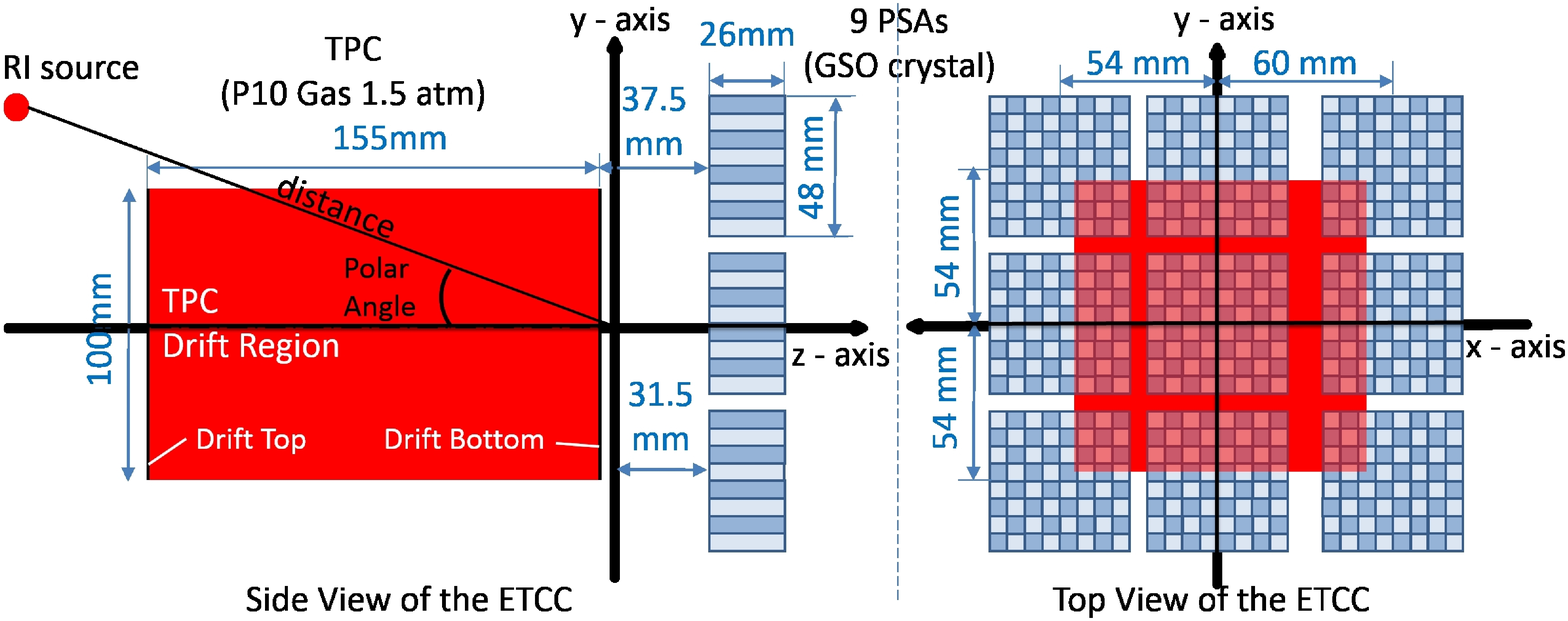}
  \caption{Schematic of an ETCC (top left). Photograph of compact and portable ETCC system for environmental gamma-ray measurements in the field (top right). Geometry of the ETCC (bottom). We define a Cartesian coordinate system as shown here. The origin of the coordinate is the center of GEM and $\mu$PIC.}
  \label{photo_schematic:ETCC}
 \end{center}
\end{figure}

To perform environmental gamma-ray measurements, we developed an ETCC that consists of a $10\times10\times16\;\rm{cm^{3}}$ TPC and a scintillation camera with 9 GSO:Ce pixelated scintillator arrays (PSAs) at the bottom of the TPC. The TPC consists of a 16$\;$cm$\;$high drift cage, 100$\;\mu$m$\;$thick gas electron multiplier (GEM) ($10\;\rm{cm}\times10\;\rm{cm}$)  \cite{gem}, and a $10\;\rm{cm}\times10\;\rm{cm}$ micro pixel chamber ($\mu$PIC). The $\mu$PIC is our original two-dimensional (2D) position-sensitive gaseous detector \cite{mu-pic2}, \cite{mu-pic3}. The TPC was set in a sealed 5$\;$mm$\;$thick aluminum vessel filled with a 1.5$\;$atm Ar-C$_{2}$H$_{6}$ gas mixture (Ar:C$_{2}$H$_{6}$ = 90:10, pressure ratio). We applied an electric field of 194$\;$V/cm between the drift top and drift bottom of the TPC, and the electron drift velocity in the TPC was approximately $4\;$cm/$\mu$sec. In the drift region, the electron cloud generated along the track of the charged particle (e.g., Compton-scattered electron) drifts along the electric field, and the electrons were multiplied by the strong electric field of the GEM and $\mu$PIC. A stable gas gain of approximately $2\times10^{4}$ was obtained in this system. The charge induced by the avalanche electrons and ions are finally detected as electrical signals by a readout circuit consisting of anode and cathode strips of the $\mu$PIC at a pitch of 800$\;\mu$m. Output signals produced on 128 anode strips and 128 cathode strips of the $\mu$PIC are amplified, shaped, and discriminated in application-specific integrated circuit (ASIC) chips on readout boards. The 128 discriminated signals in the ASIC chips are fed to a field-programmable gate array (FPGA) on the same board and the hit signals are synchronized with the 100$\;$MHz clock. The position and elapsed time from the trigger are also recorded. However, the analog signals are summed via the ASIC chips into four-channels, and their waveforms are recorded as eight-bit and 25-MHz data by four-channel flash-ADC on the board. The FPGA controls the data acquisition of the hit and digital waveform data of the TPC. The details are given in reference \cite{mizumoto-ieee}.

We use a 64-channel multi-anode position-sensitive photomultiplier tube (PMT, Hamamatsu Photonics H8500) for the photodetector of the PSA. The PMT is $52\times52\;\rm{mm}^{2}$ (28$\;$mm high), and its active photo-cathode area is $49\times49\;\rm{mm}^{2}$ with a matrix of $8\;\times8$ anodes. For a high-voltage supply to the PMTs, we used a dedicated compact high-voltage power supply board developed for the SMILE-II FM ETCC \cite{mizumoto-ieee}. Each GSO:Ce (Gd$_{2}$SiO$_{5}$:Ce) PSA consists of $8\times8$ pixels. Each pixel crystal is $6\;\times\;6\;\rm{mm}^{2}$ ($26\;\rm{mm}$ thick); this pitch is the same as that of the anode of the H8500. Each pixel is optically isolated by a Vikuiti 3M enhanced specular reflector, and each PSA is glued to the H8500 with SAINT-GOBAIN BC-600 optical cement. We constructed a scintillation camera with $3\times3$ PSAs as shown in the bottom panel of figure \ref{photo_schematic:ETCC}. 

Using the TPC and scintillation camera as described above, we constructed an ETCC dedicated for environmental gamma-ray measurements. For outdoor measurements, we developed a system that consists mainly of the ETCC, VMEs, DAQ control boards, dc/dc converters, and a rechargeable lithium-polymer battery; all are attached to a frame as shown in the top-right photograph of figure \ref{photo_schematic:ETCC}.  The weight and power consumption of the device are approximately 100$\;$kg and 100$\;$W, respectively. Because the charge capacity of the battery is approximately 25$\;$V 100$\;$Ah, this system can be operated for a day without an external power supply. The height of the system is approximately 1.7$\;$m.

\section{Gamma-ray reconstruction}
\label{sect:image}
\begin{figure}[t]
 \centering
 \includegraphics[width=7.cm]{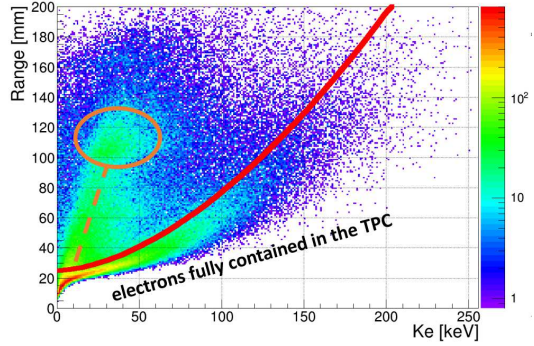}
 \caption{Track range vs. energy deposit (dE/dx) of charged particles in the TPC under $^{137}$Cs gamma-ray irradiation. The events caused by minimum-ionization particles (MIPs) such as cosmic muons that enter from one face of the TPC and go out the opposite face are in the region around the orange-colored circle. The MIP events that slightly enter the TPC volume or electrons that are generated in the TPC and leave the TPC are distributed near the orange-colored dashed line. The solid red line indicates the dE/dx cut criteria for the discrimination of fully-contained electrons from other charged particles in this analysis.}
 \label{fig:dedx}
\end{figure}

\begin{figure}[t]
 \begin{center}  
  \begin{tabular}{c}
   \begin{minipage}{0.33\hsize}
    \begin{center}
     \includegraphics[width=4.8cm]{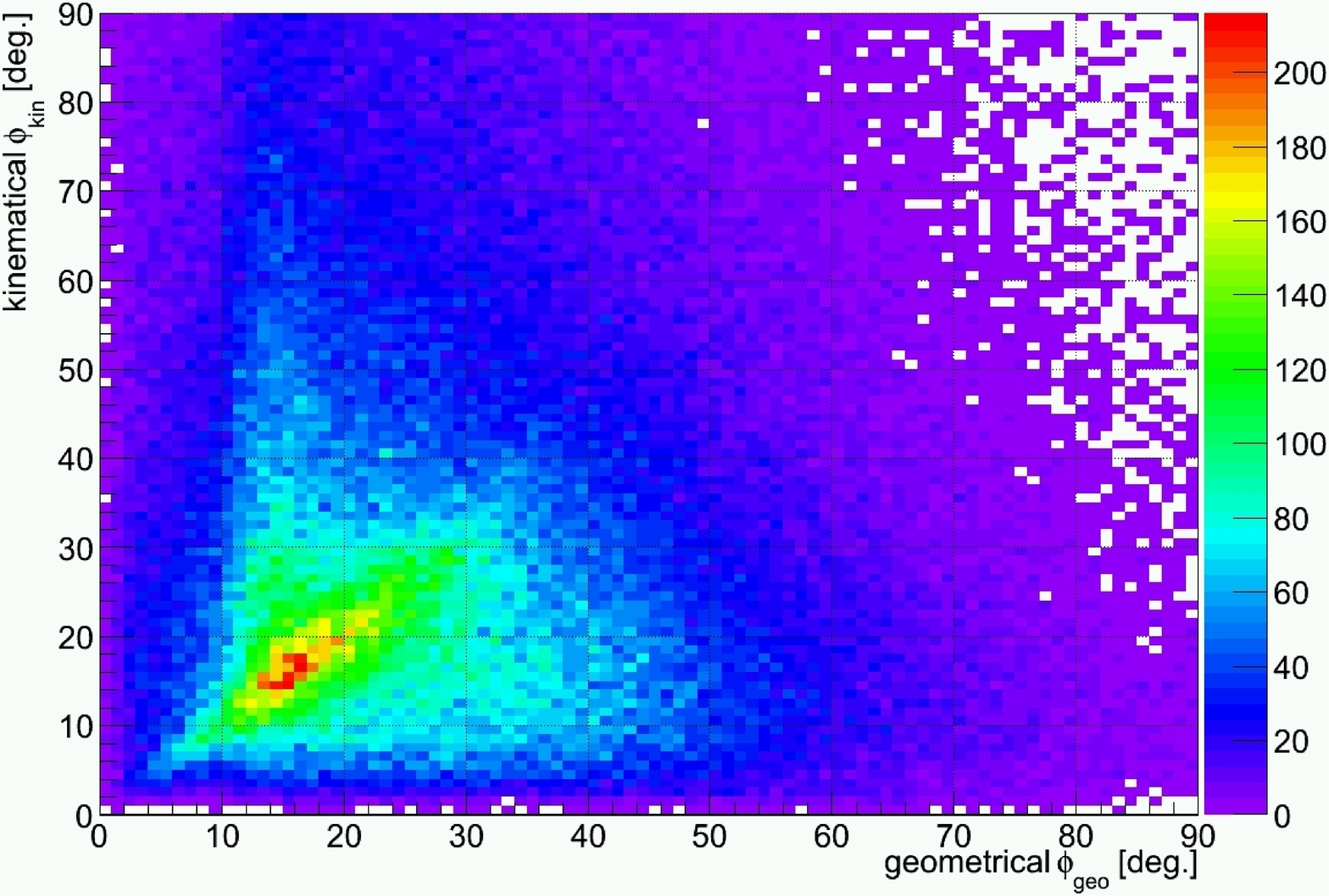}
    \end{center}
   \end{minipage}
   \begin{minipage}{0.33\hsize}
    \begin{center}
     \includegraphics[width=4.8cm]{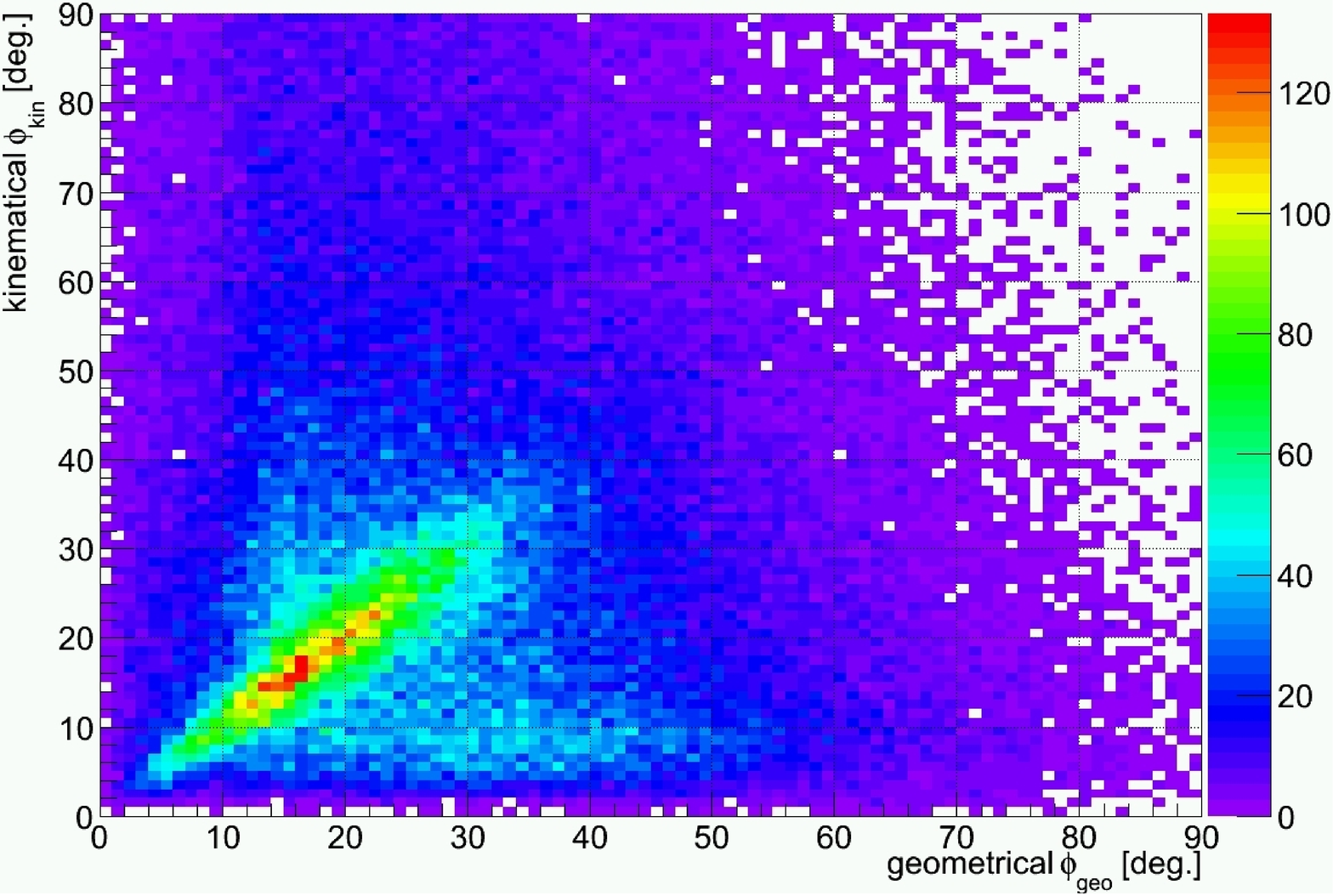}
    \end{center}
   \end{minipage}
   \begin{minipage}{0.33\hsize}
    \begin{center}
     \includegraphics[width=4.8cm]{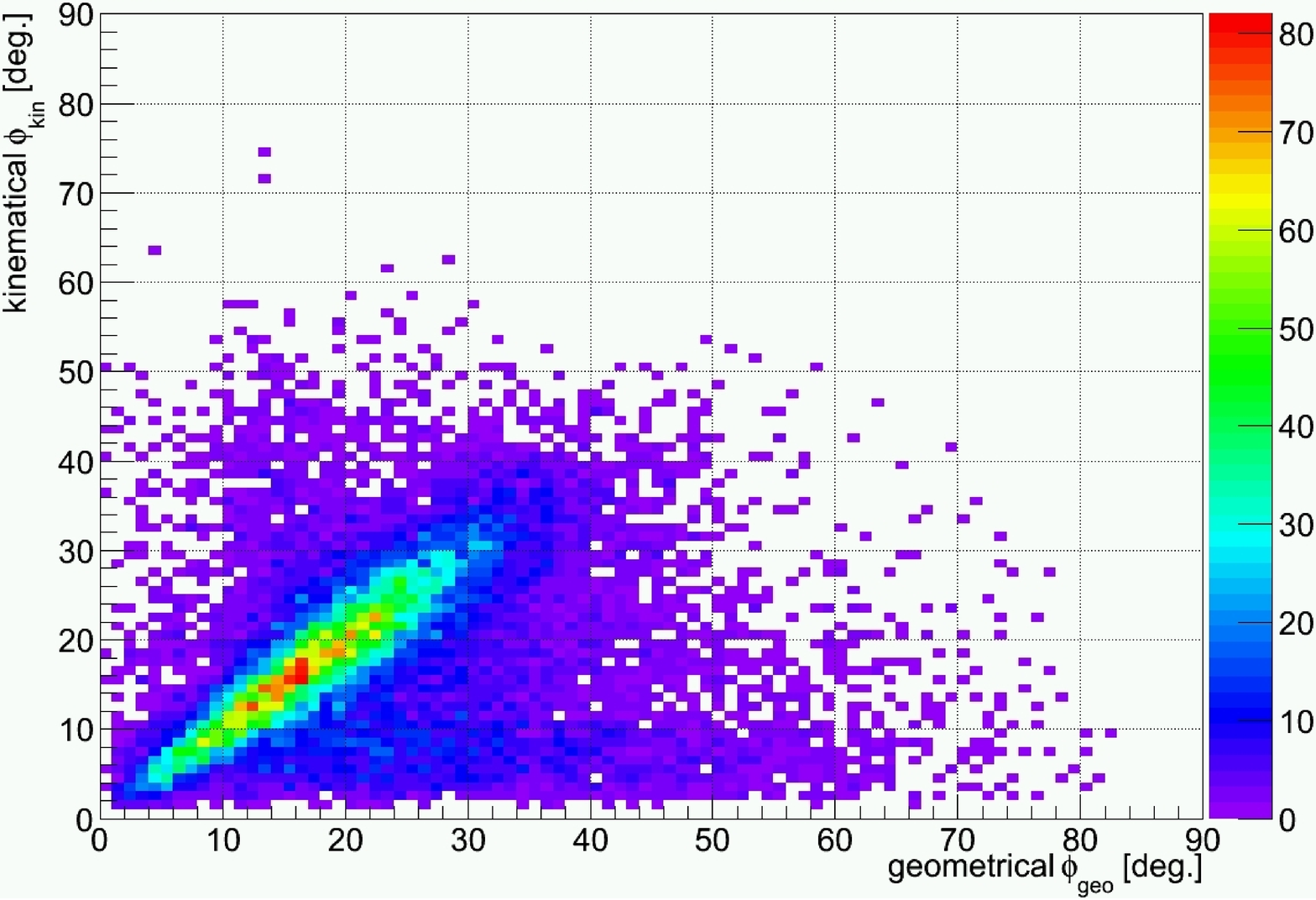}
    \end{center}
   \end{minipage}
  \end{tabular}
  \caption{Relationship between kinematically determined Compton scattering angle $\phi_{\rm{kin}}$ and geometrically determined angle $\phi_{\rm{geo}}$ for 662$\;$keV gamma-ray irradiation. The left and center panels show the $\phi_{\rm{kin}}-\phi_{\rm{geo}}$ relationship before and after the dE/dx selection, respectively. The right panel includes volume and energy selections after the dE/dx selection. The details of the selections are described in the text.}
  \label{fig:phi}
 \end{center}
\end{figure}

\subsection{Criteria for gamma-ray selection from source}
To reconstruct both the direction and energy of the incident gamma ray, we need to know the Compton-scattering vertex and the recoil direction of the Compton electron, which we can obtain by analyzing the x-z and y-z electron-track data from the TPC. In the analysis described in sections 3 and 4, we use the electron-track analysis used in reference \cite{mizumura_SMILE-II}. We have been studying the improved electron tracking analyzing method.

In the analysis, we use three criteria to select gamma rays from the source. As the first criterion, we use the energy-loss rate (dE/dx) information of charged particles in the TPC. Figure \ref{fig:dedx} shows the measured dE/dx map of the ETCC. This map shows that we can clearly separate two components above and below the solid red line whose relational expression is
\begin{eqnarray}
\left(\frac{\rm{Track\;Range}}{\rm{[mm]}}\right) = 2.7 \times 10^{3} \left( \frac{K_{e}}{\rm[MeV]}\right)^{1.72} + 25,
\end{eqnarray}
where $K_{e}$ is the energy deposited by a charged particle in the TPC, and the track range is determined by the geometrical combination of the two 2D tracks. The background events above the line are MIP-like events (where "MIP" refers to minimum ionizing particles), such as high-energy electrons escaping from the TPC and cosmic muons. The component below the line is from fully-contained electrons in the TPC. In the analysis, we use the high dE/dx component below the line. As the second criterion, we use the fiducial volume cut. We select events whose scattering vertex is in the region $-50\;\rm{mm}\;\leq\;x, y\;\leq\;+50\;\rm{mm}$, and $-161\;\rm{mm}\;\leq\;z\;\leq\;-6\;\rm{mm}$, which is the full drift region (detection area) of the TPC (cf. figure \ref{photo_schematic:ETCC}). As the third criterion, we use the energy selection of the ETCC. We use twice the energy-resolution range (full width at half maximum (FWHM)) of the energy-selection window. From the result of measuring $^{137}$Cs, $^{22}$Na, and $^{54}$Mn at the center of the FoV, we obtain the energy resolutions (FWHM) of $12.7\% \pm 1.3\%$ at 511$\;$keV, $11.6\% \pm 1.3\%$ at 662$\;$keV, $10.2\% \pm 1.9\%$ at 835$\;$keV, and $8.4\% \pm 1.9\%$ at 1275$\;$keV, and the best fit of these values is $\Delta\rm{E/E\;(FWHM)} = (11.4\pm0.8) \times (\rm{energy}/662\;\rm{[keV]})^{-0.45\pm0.24}\;[\%]$. The energy resolution of the ETCC is mainly affected by that of the PSA.

Figure \ref{fig:phi} shows the measured relationship between the kinematically determined Compton-scattering angle $\phi_{\rm{kin}}$ and the geometrically calculated scattering angle $\phi_{\rm{geo}}$ under 662$\;$keV gamma-ray irradiation.  The angle $\phi_{\rm{kin}}$ is determined from the total energy and the ratio of energies of the recoil electron and scattered gamma ray. The angle $\phi_{\rm{geo}}$ is determined from the recoil direction and Compton-scattering position with the assumption of the gamma-ray-source direction. From the figure, backgrounds are suppressed, and we can observe a strong correlation between $\phi_{\rm{kin}}$ and $\phi_{\rm{geo}}$, which shows the detection of source-derived gamma rays.

\subsection{Gamma-ray imaging}

\begin{figure}[t]
 \begin{center}  
  \begin{tabular}{c}
   \begin{minipage}{0.25\hsize}
    \begin{center}
     \includegraphics[width=3.4cm]{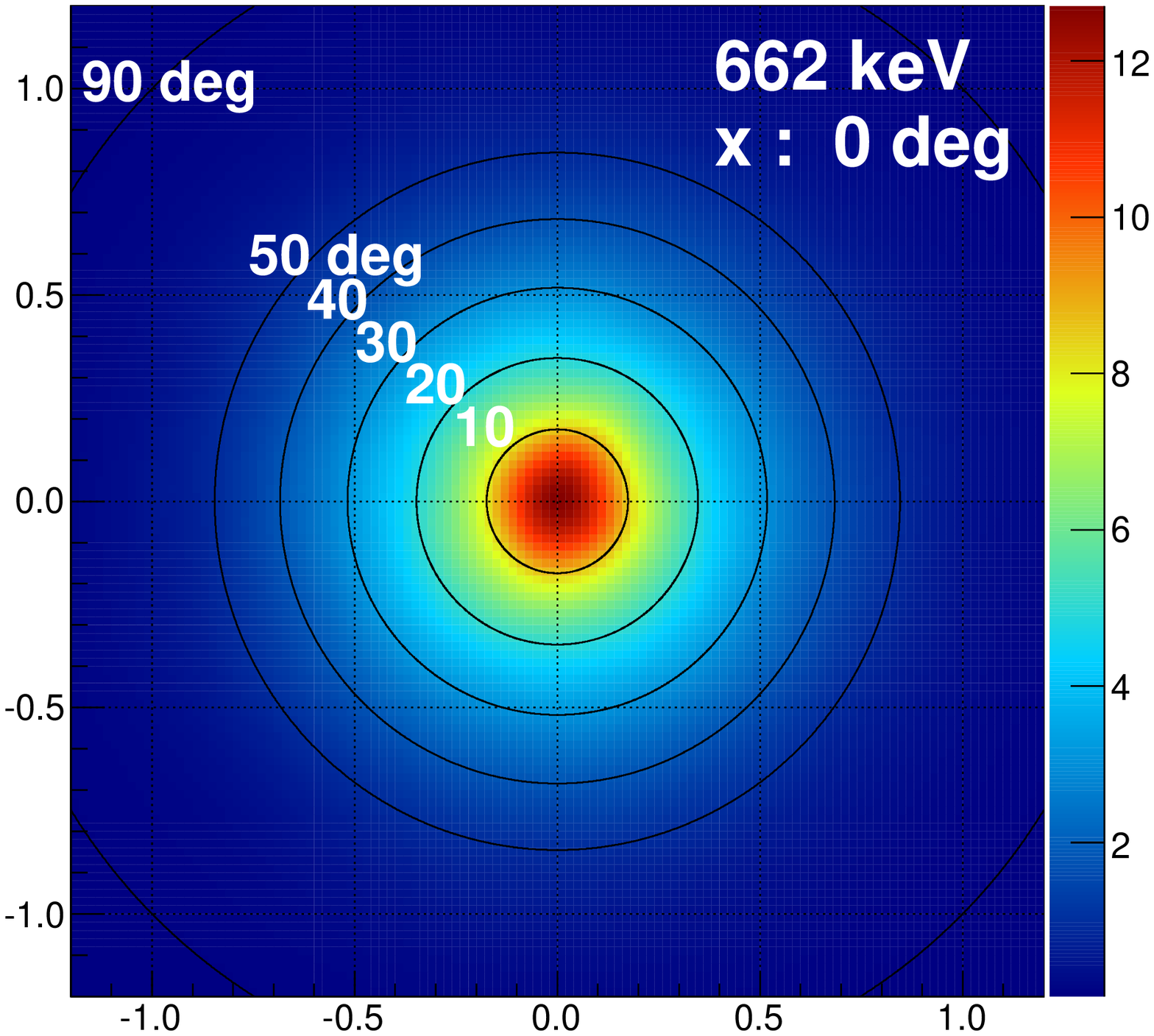}
    \end{center}
   \end{minipage}
   \begin{minipage}{0.25\hsize}
    \begin{center}
    \includegraphics[width=3.4cm]{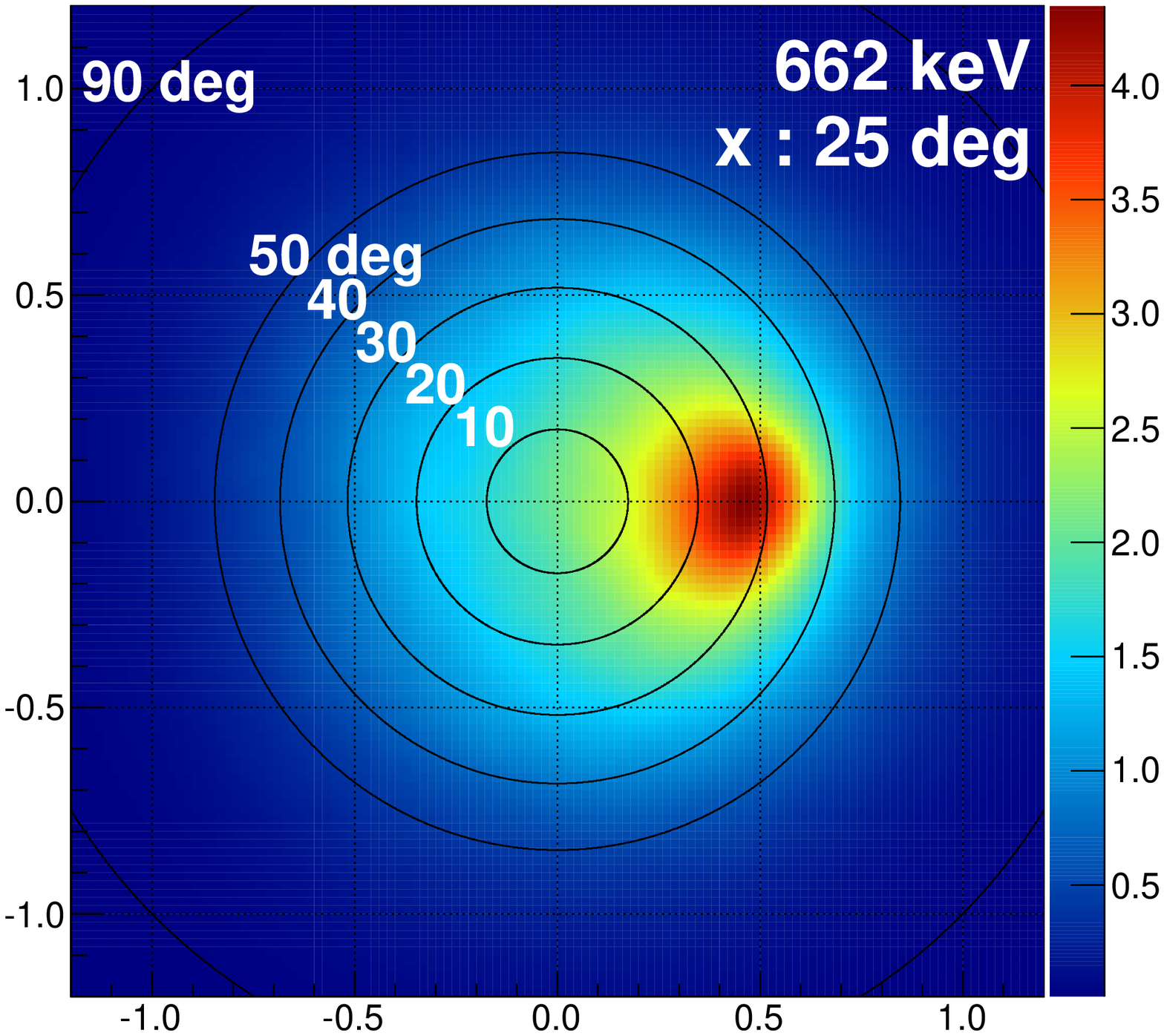}
    \end{center}
   \end{minipage}
   \begin{minipage}{0.25\hsize}
    \begin{center}
    \includegraphics[width=3.4cm]{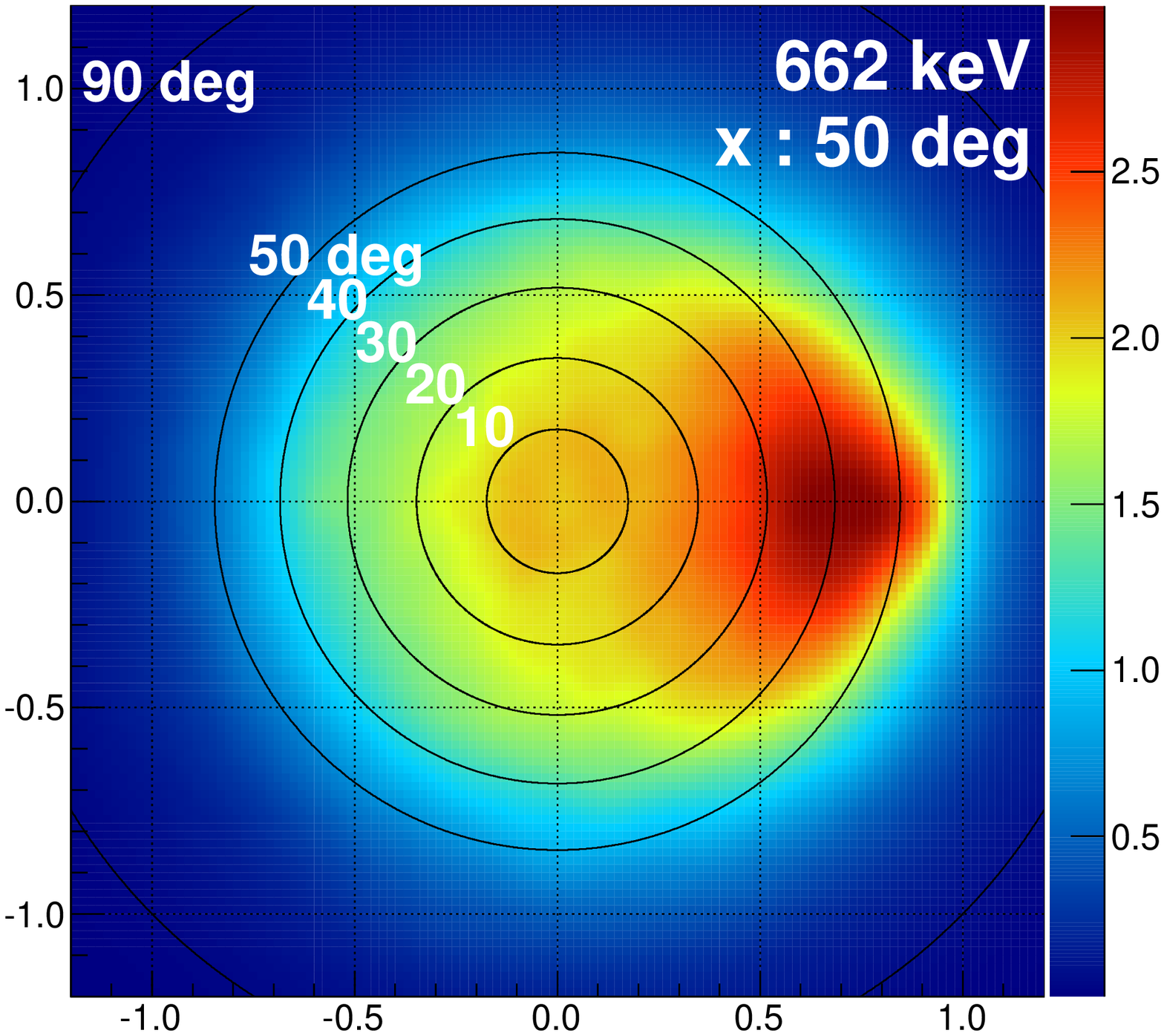}
    \end{center}
   \end{minipage}
   \begin{minipage}{0.25\hsize}
    \begin{center}
    \includegraphics[width=3.4cm]{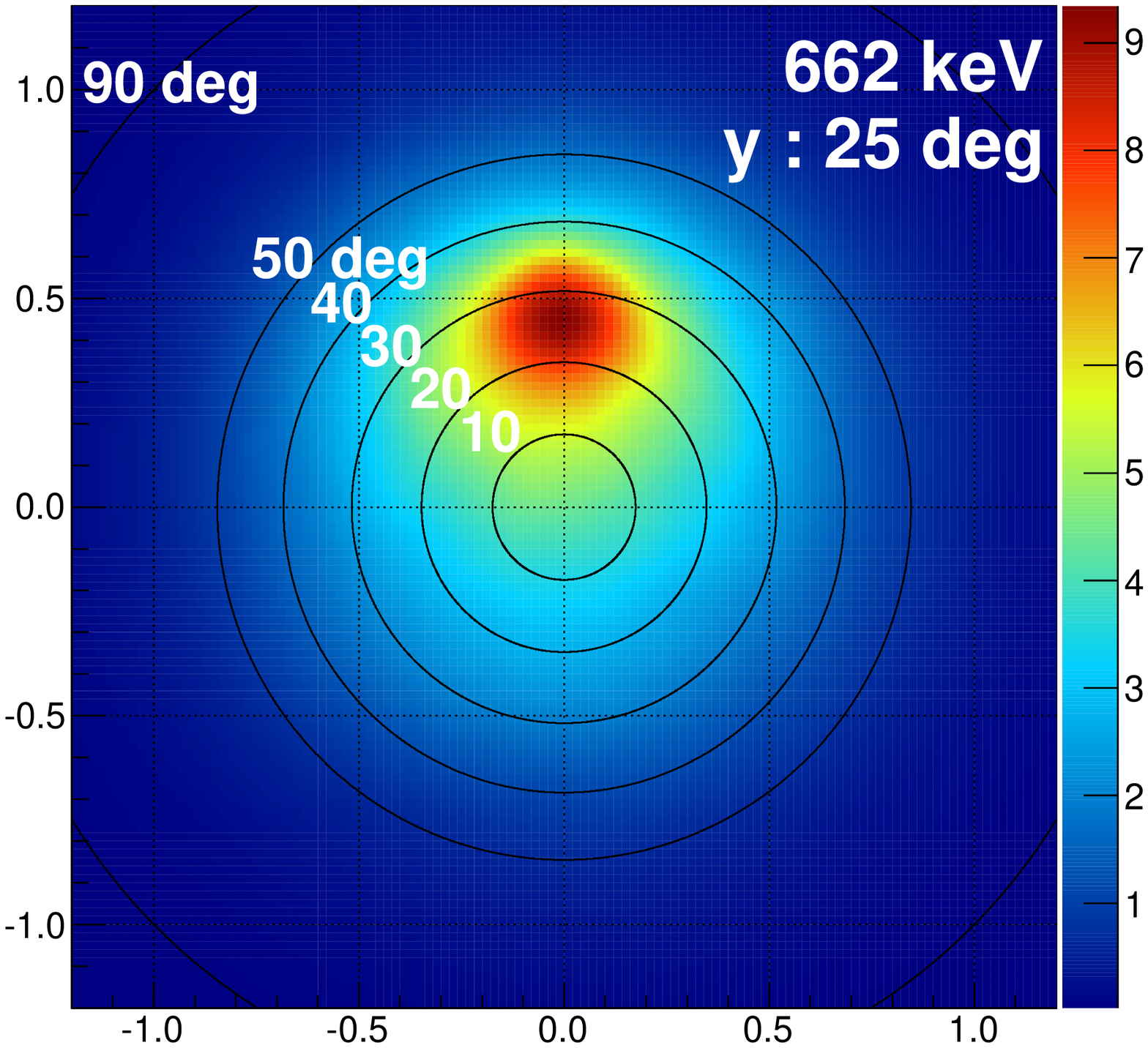}
    \end{center}
   \end{minipage}
  \end{tabular}
  \caption{Demonstration of gamma-ray images from $^{137}$Cs gamma-ray source. These images are drawn in the Lambert azimuthal equal-area projection. We set a $^{137}$Cs point source 50$\;$cm from the origin of the coordinate described in figure 1 and the polar angles of 0$^{\circ}$, +25$^{\circ}$ and +50$^{\circ}$ in the x direction and +25$^{\circ}$ in the y direction from the left panel to right panel, respectively. In drawing this figure, we did not correct the acceptance of the ETCC. }
  \label{fig:gamma-image}
 \end{center}
\end{figure}

Figure \ref{fig:gamma-image} shows gamma-ray images of the $^{137}$Cs gamma-ray point source. From left to right, we set the source at polar angles of 0$^{\circ}$, 25$^{\circ}$, and 50$^{\circ}$ along the x axis and 25$^{\circ}$ along the y axis. If we set the source at the polar angles of 0$^{\circ}$, 25$^{\circ}$, or 50$^{\circ}$, the source images can be observed on the points whose distances from the center of the images are 0, 0.43, and 0.85, respectively. From these figures, we see the change of the source position in the region of the polar angles from $0^{\circ}$ to $50^{\circ}$. We also see that the source image becomes unclear if the polar angle becomes larger. One reason is the worse energy resolution. The larger is the Compton scattering angle, the smaller is the energy of the Compton-scattered gamma ray. The small absorption energy at the PSA leads to poor energy resolution \cite{ueno_jinst}. Another reason is the acceptance distribution of the ETCC. As will be discussed in the next section, the efficiency at a small polar angle is better than that at a large polar angle. If we correct the acceptance of the ETCC, we obtain a more point-like image.

\section{Performance of the ETCC}
\label{sect:performance}
\begin{figure}[t]
 \begin{center}  
  \begin{tabular}{c}
   \begin{minipage}{0.5\hsize}
    \begin{center}
     \includegraphics[width=7.3cm]{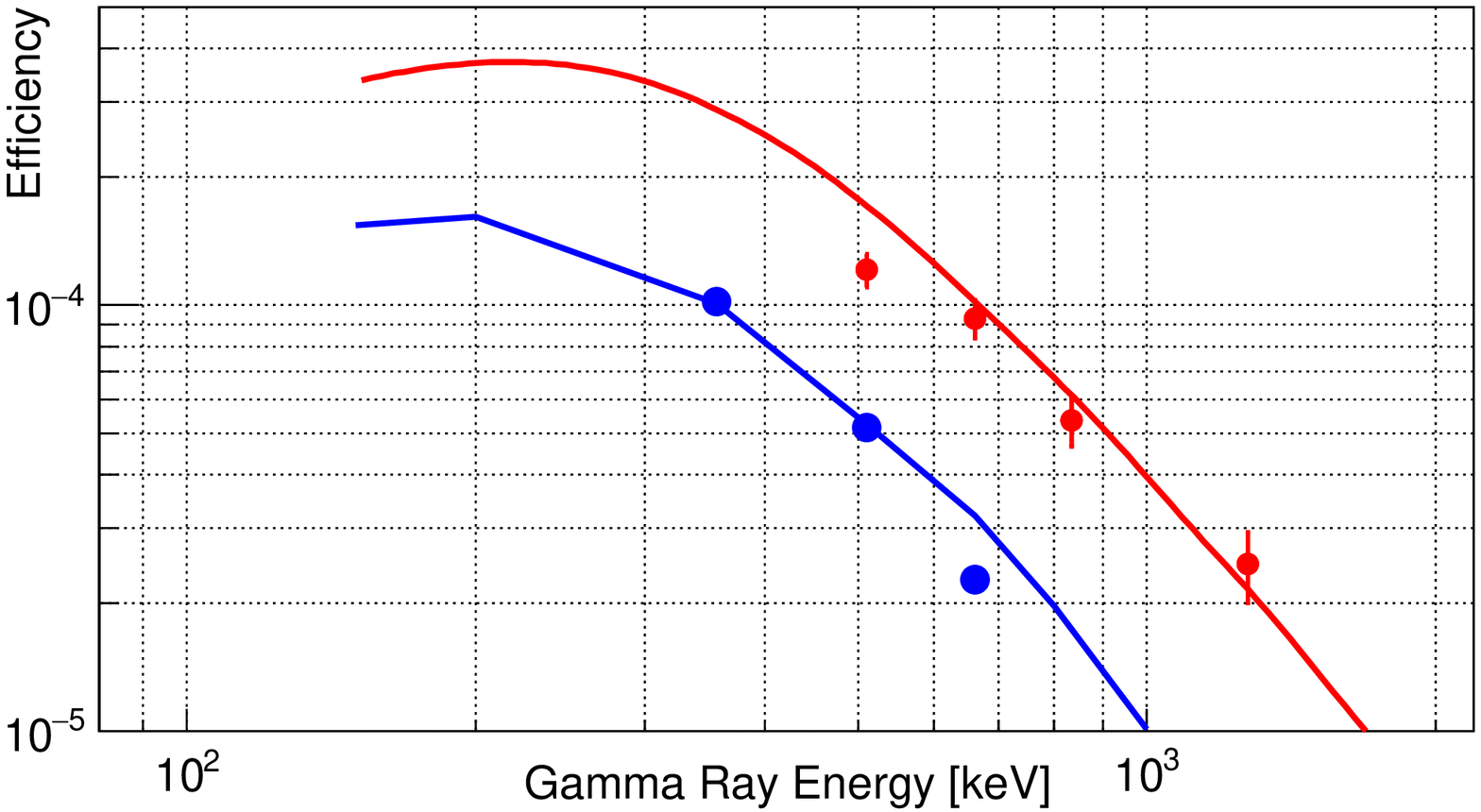}
     \hspace{1.6cm}
    \end{center}
   \end{minipage}
   \begin{minipage}{0.5\hsize}
    \begin{center}
     \includegraphics[width=7.3cm]{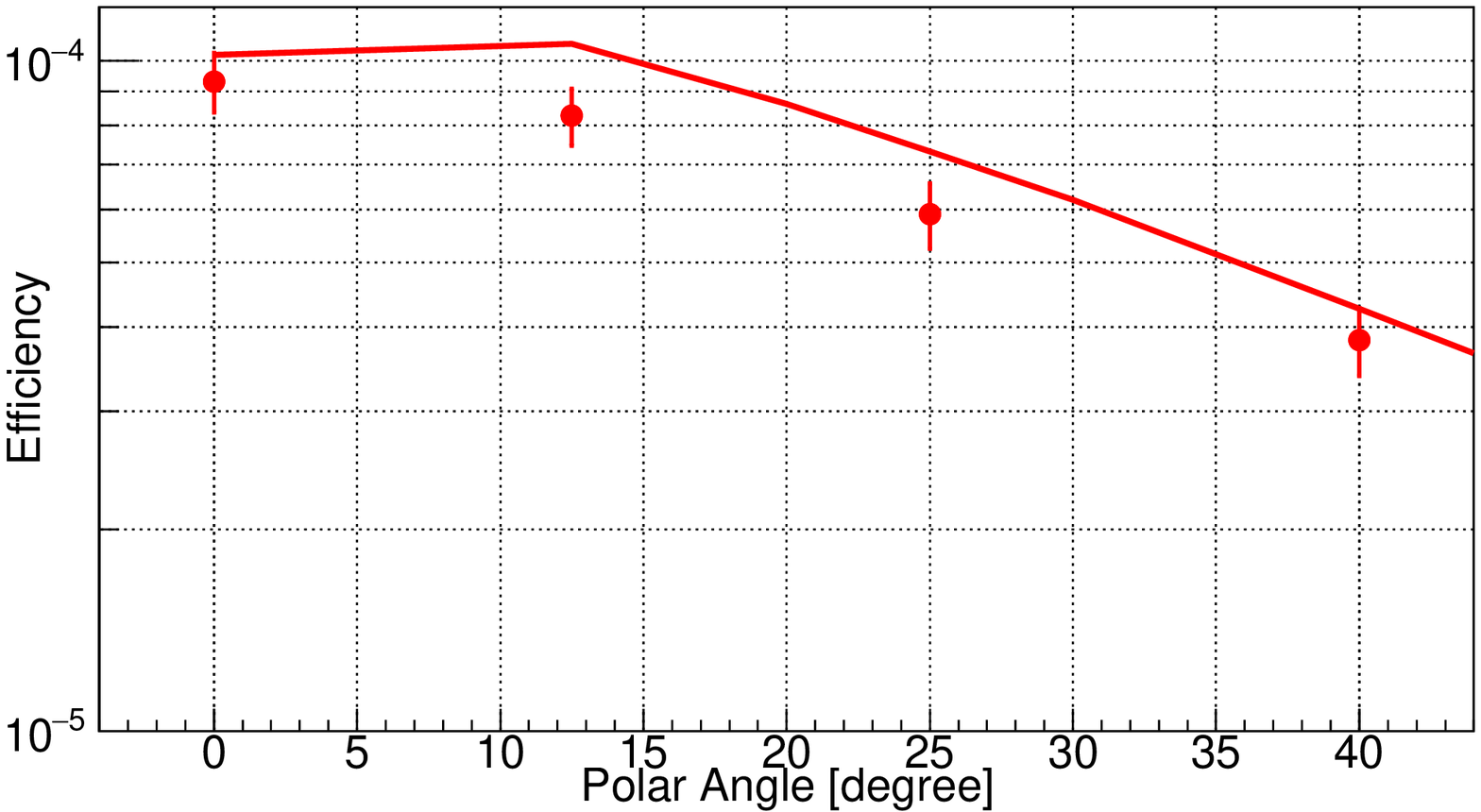}
     \hspace{1.6cm}
    \end{center}
   \end{minipage}
  \end{tabular}
  \caption{(Left): Red points and line show the measured and simulated results of the energy dependence of the detection efficiency of the ETCC for gamma rays from the center of the FoV. The blue points and line are the measured and simulated results for a similar ETCC that used 1-atm gas for the TPC and GSO pixels of thickness 13$\;$mm, respectively \cite{komura-ieee},\cite{sawano_simulation}. (Right): Red points and line show the measured and simulated data of the angular dependence of the 662$\;$keV gamma-ray detection efficiency of the ETCC, respectively.}
  \label{plot:efficiency}
 \end{center}
\end{figure}

In this section, we show the results of performance tests based on laboratory exposures to point gamma-ray sources $^{22}$Na (511$\;$keV, 1275$\;$keV), $^{137}$Cs (662$\;$keV), and $^{54}$Mn (835$\;$keV). 

\subsection{Gamma-ray detection efficiency}
We measured the energy and angular dependence of the gamma-ray detection efficiency of the ETCC. To calculate the detection efficiency, we must know the total number of gamma rays that enter the fiducial volume of the TPC during the measurement. We calculated this number from the intensity and position of the gamma-ray source. We also derive the number of detected gamma rays by using all cut conditions (dE/dx cut, fiducial volume cut, and energy cut) described in section 3.1 and the subtraction of the room background events from the background measurement data. We show the measured and simulated results in figure \ref{plot:efficiency}. For the calculation of the red lines, we used Geant4 \cite{geant4} and only considered the 1.5$\;$atm Ar-C$_{2}$H$_{6}$ gas mixture in the drift region of the TPC and GSO scintillator arrays and do not consider other structure such as the vessel. For the calculation of the efficiency from the simulated result, we followed the event detection criteria written in reference \cite{sawano_simulation}. As shown in the figure, the detection efficiency for 662$\;$keV gamma rays from the center of the FoV is $(9.31 \pm 0.95) \times 10^{^-5}$. It is approximately four times higher than the result of another ETCC, mainly because the gas pressure was 1.5 times higher and the scintillators were twice as thick. The measured and simulated results are consistent to within 42$\%$ in 511-1275$\;$keV; this shows that the ETCC detects almost all Compton events and discriminates well against background events. If we define a FoV as the angle that encompasses half of the detection efficiency at its center, then the FoV is approximately 1$\;$sr.

\subsection{Angular resolution measure}

\begin{figure}[t]
 \begin{center}  
  \begin{tabular}{c}
   \begin{minipage}{0.5\hsize}
    \begin{center}
     \includegraphics[width=7.3cm]{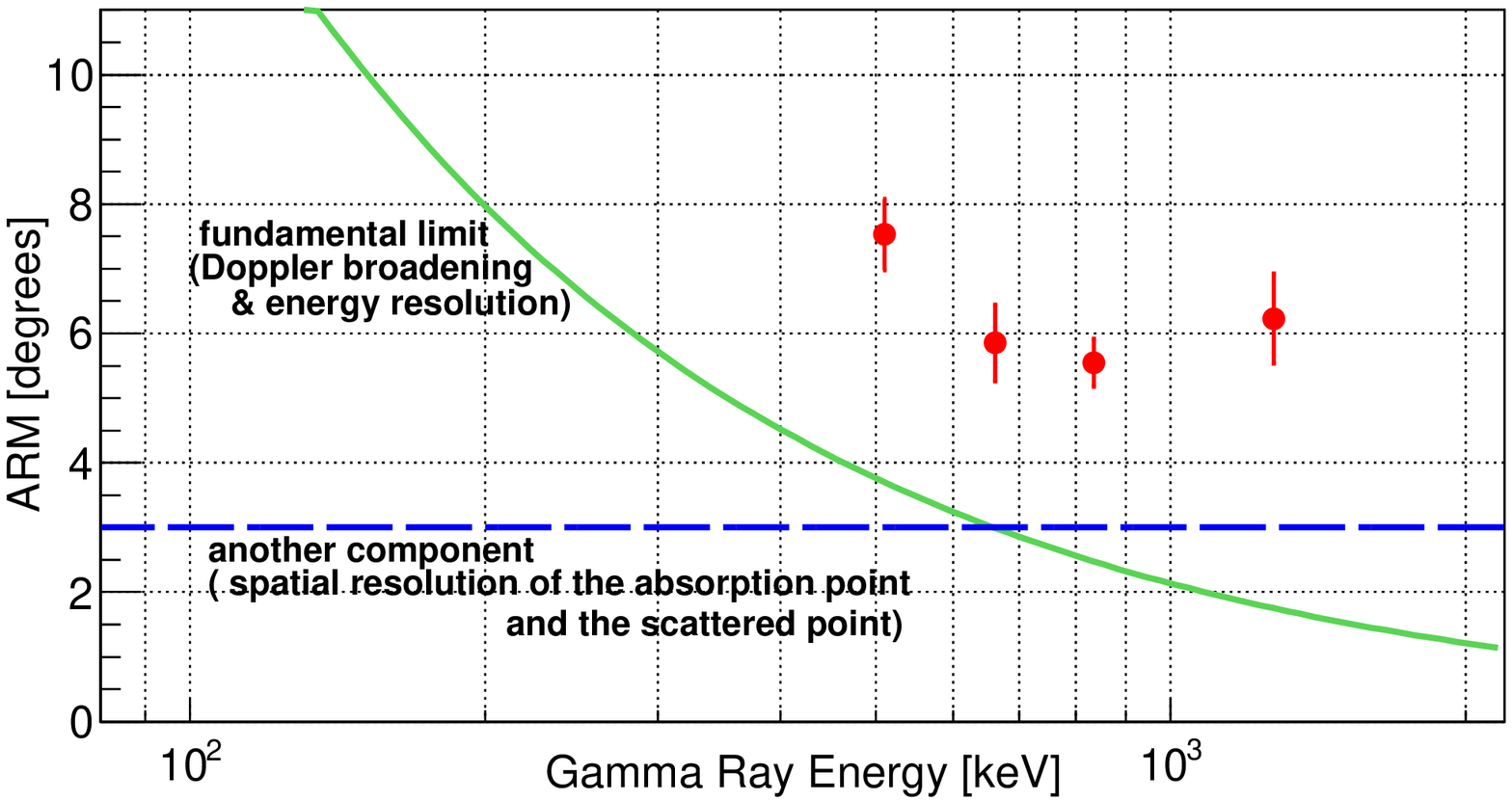}
     \hspace{1.6cm}
    \end{center}
   \end{minipage}
   \begin{minipage}{0.5\hsize}
    \begin{center}
     \includegraphics[width=7.3cm]{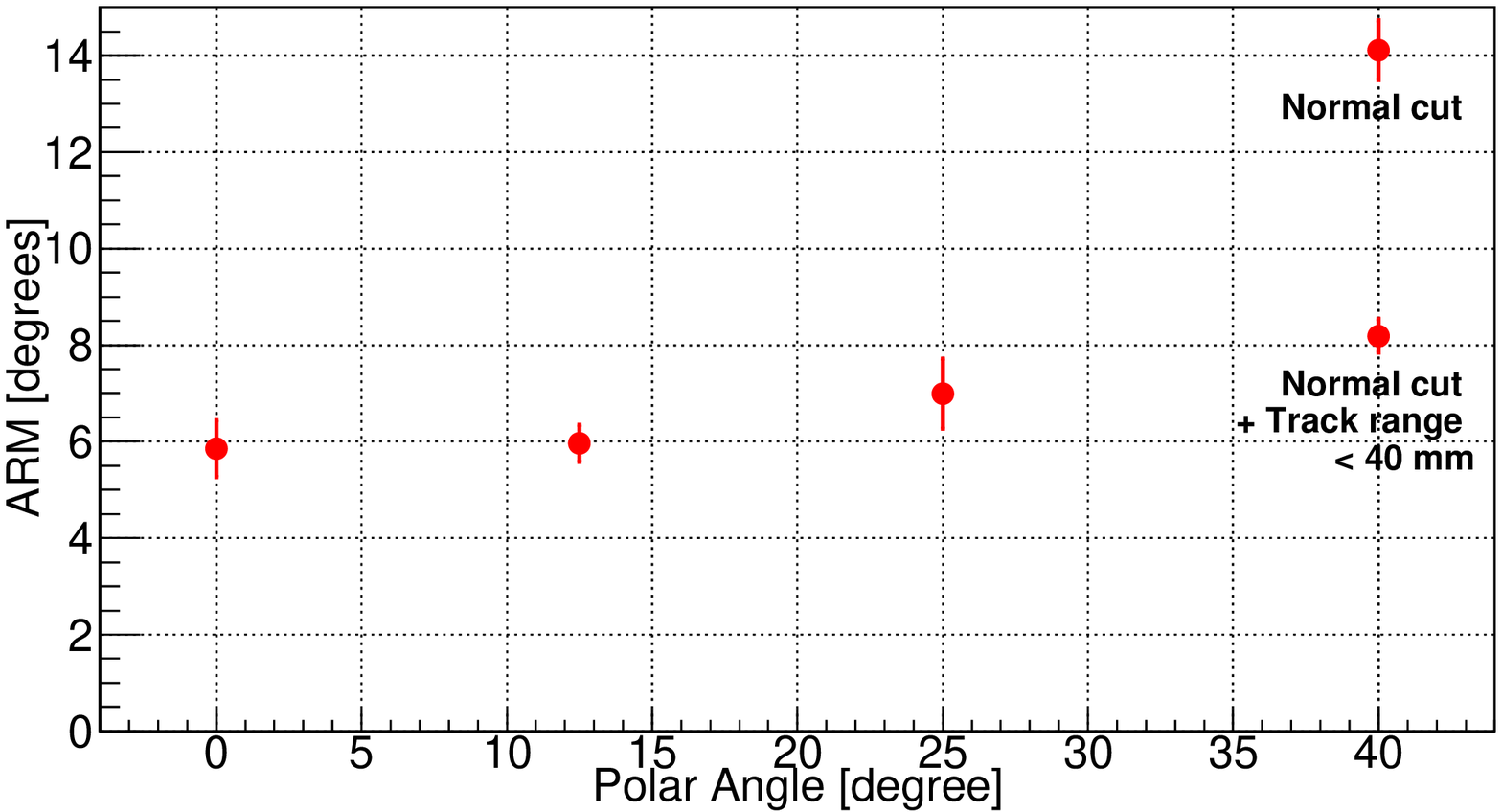}
     \hspace{1.6cm}
    \end{center}
   \end{minipage}
  \end{tabular}
  \caption{(Left): The measured results of the energy dependence of the ARM (FWHM) for gamma rays from the center of the FoV. The green solid line indicates the fundamental limit of the ARM in consideration of Doppler broadening and the energy resolution of TPC and GSO. The blue dashed line shows the gap between the measured result and green line. (Right): the measured results of the angular dependence of the ARM for 662$\;$keV gamma rays. At the polar angle of 40$^{\circ}$, we fitted the ARM distribution with a Gaussian and we plot two points with different cut conditions.}
  \label{plot:arm}
 \end{center}
\end{figure}

We also checked the ARM of the ETCC from the measured data. The ARM is defined by $\phi_{\rm{geo}}$ - $\phi_{\rm{kin}}$. For the ARM calculation, we use the dE/dx cut, the fiducial volume cut, and the energy cut described in section 3.1, and we fit the ARM distribution with a Lorentzian function unless otherwise noted. Figure \ref{plot:arm} shows the measured ARM energy dependence (FWHM) for gamma rays from the center of the FoV (left panel) and the polar angular dependence of the ARM (FWHM) for 662$\;$keV gamma rays. From the left panel of this figure, we observe that the ARM for 662$\;$keV gamma rays from the center of the FoV is $5.9^{\circ} \pm 0.6^{\circ}$. In the left panel of the figure, we show the result for the fundamental limit obtained from the simulation when considering only the energy resolution and Doppler broadening. We observe that the gap between the measured data and green line is approximately 3$^{\circ}$ or 4$^{\circ}$ from 511 to 1275$\;$keV. The gap is mainly because of the spatial accuracy with which the absorption and scattered points are determined. From the right panel of the figure, we observe that the ARM becomes worse at high polar angles because of the worse energy resolution, as mentioned in section 3.2. We can prevent the deterioration of the ARM at high polar angles by optimizing the number and position of PSAs.

\begin{figure}[t]
 \begin{center}  
  \begin{tabular}{c}
   \begin{minipage}{0.247\hsize}
    \begin{center}
     \includegraphics[width=3.1cm]{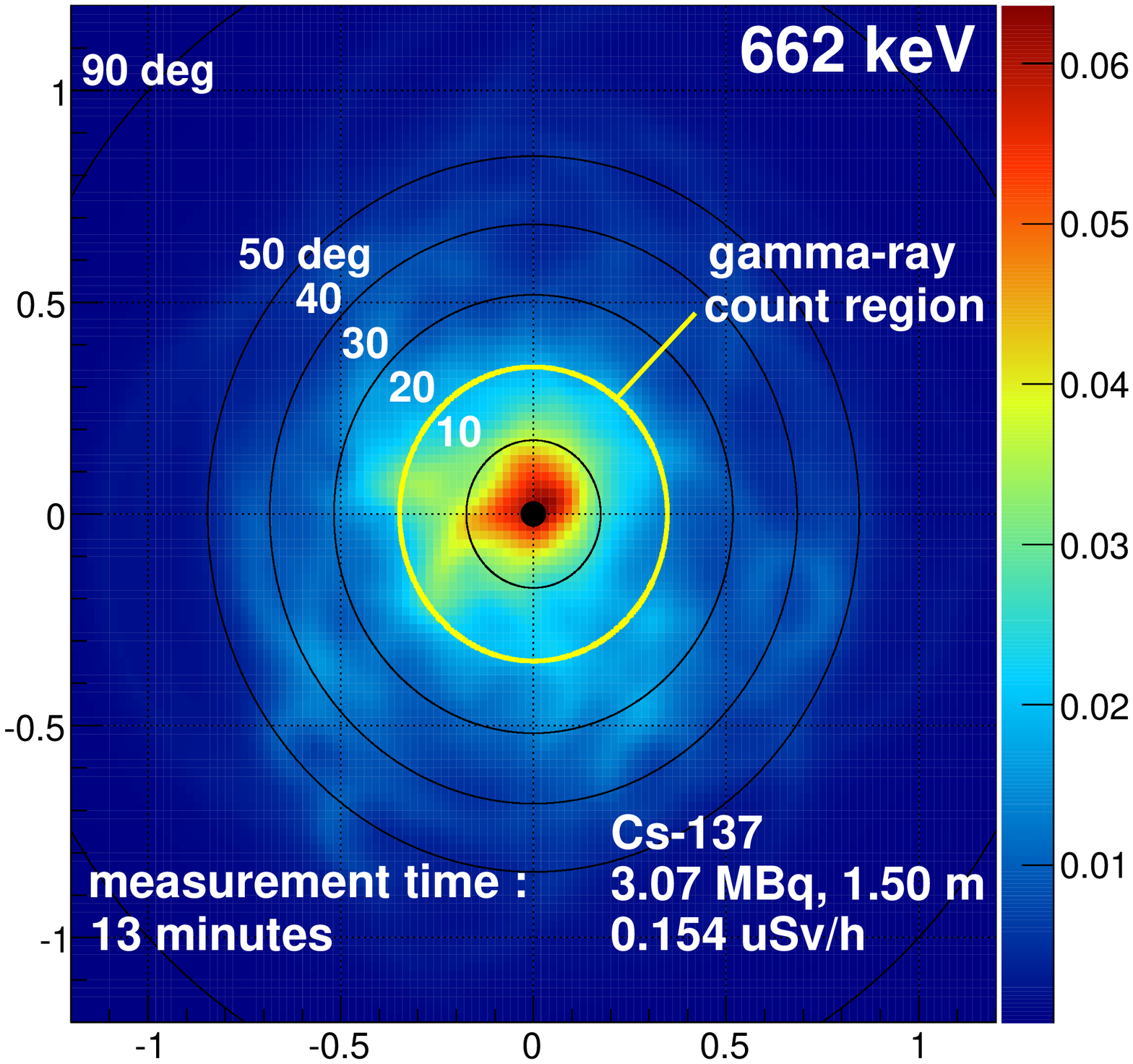}
    \end{center}
   \end{minipage}
   \begin{minipage}{0.251\hsize}
    \begin{center}
     \includegraphics[width=3.3cm]{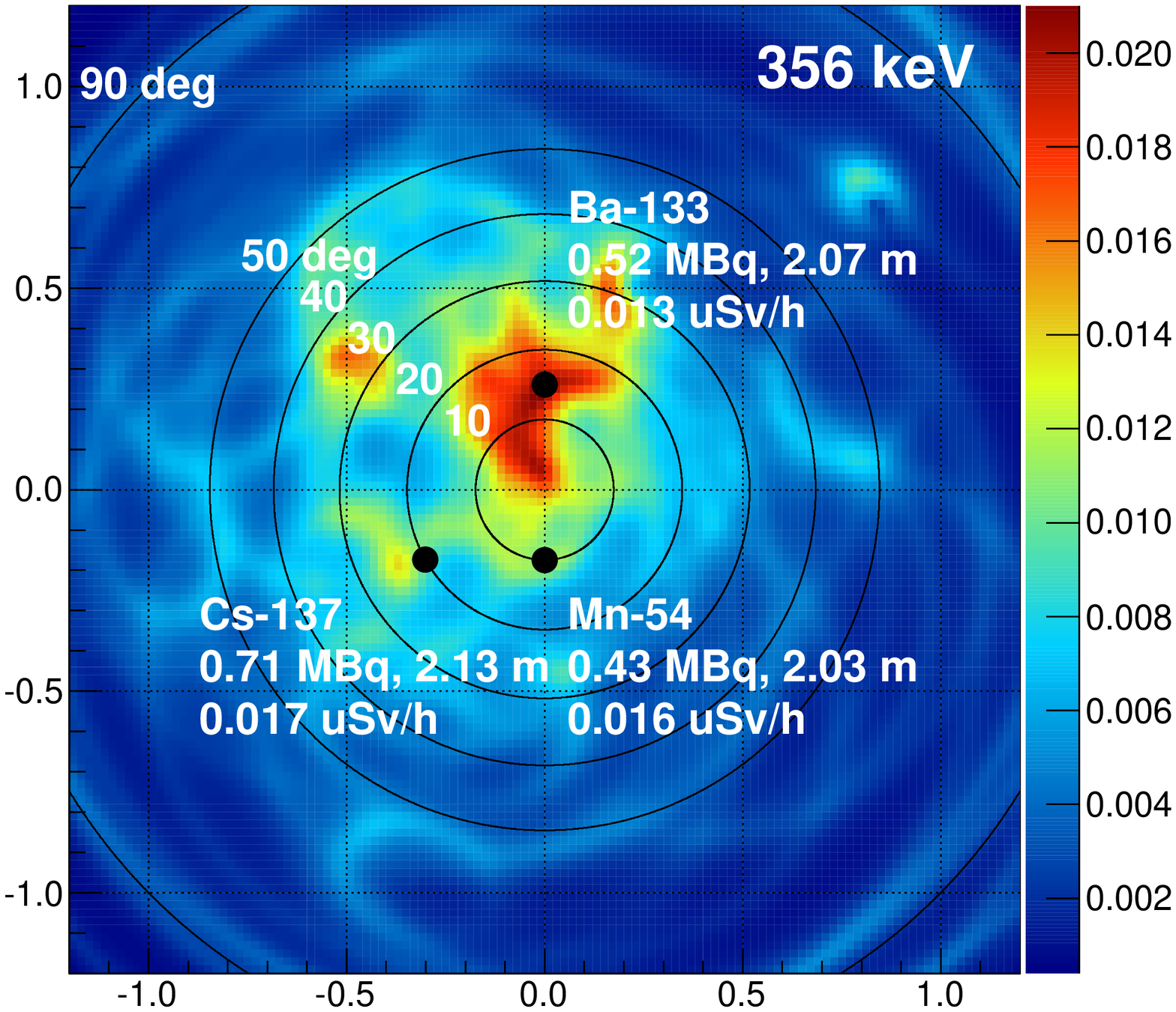}
    \end{center}
   \end{minipage}
   \begin{minipage}{0.251\hsize}
    \begin{center}
    \includegraphics[width=3.3cm]{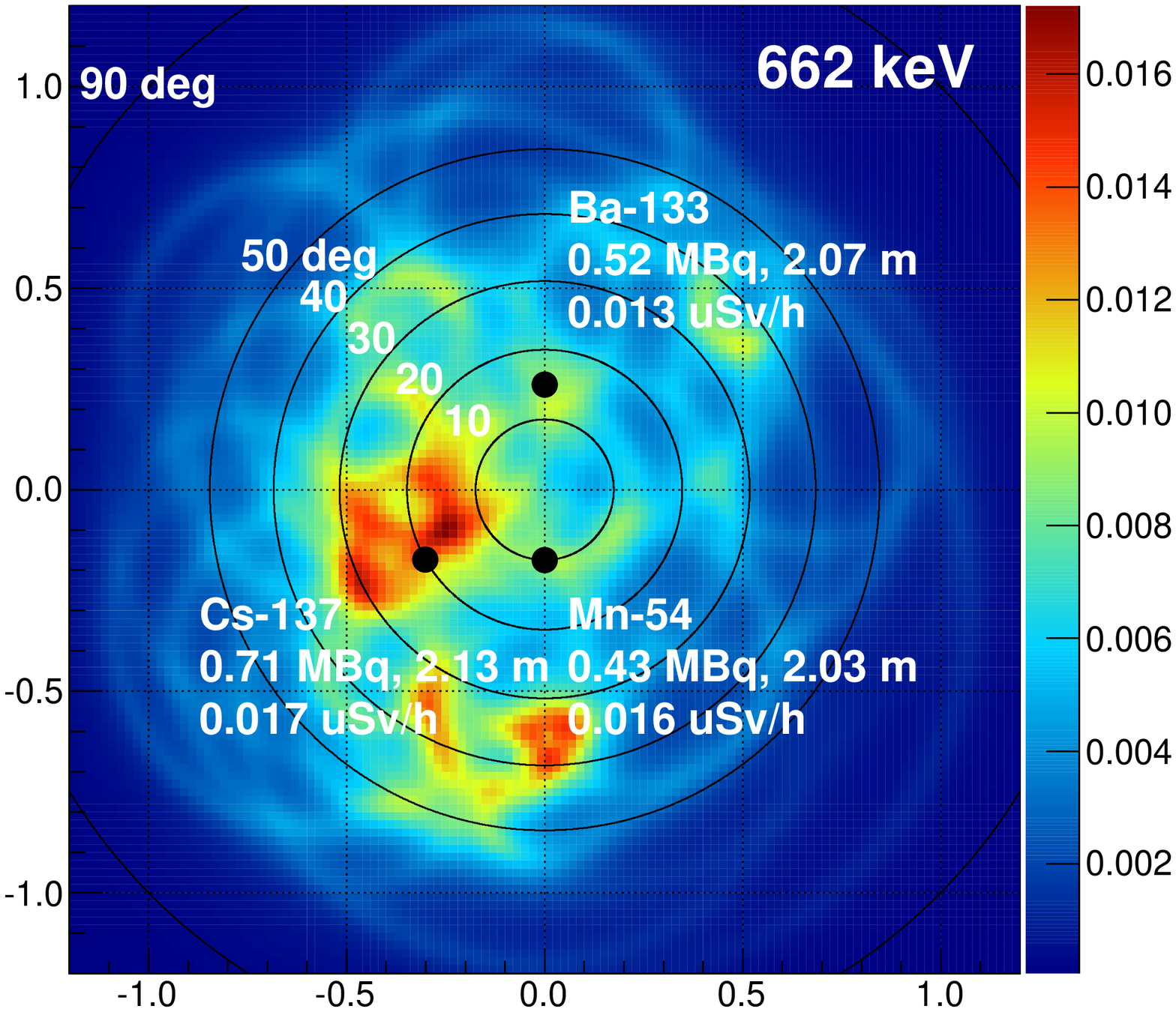}
    \end{center}
   \end{minipage}
   \begin{minipage}{0.251\hsize}
    \begin{center}
      \includegraphics[width=3.3cm]{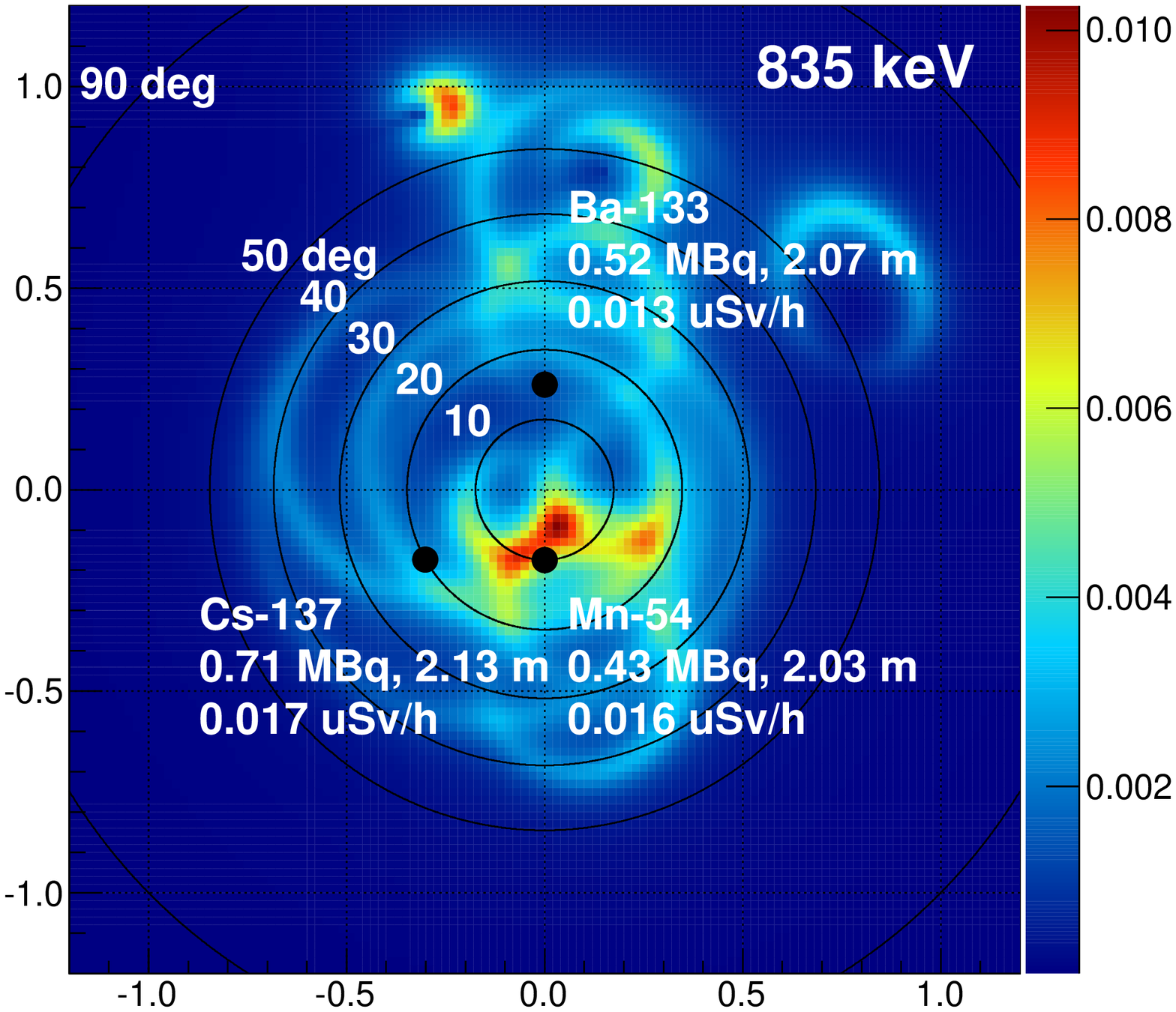}
    \end{center}
   \end{minipage}
  \end{tabular}
  \caption{Left panel shows the 5$\sigma$ detection image of 662$\;$keV gamma rays from a $^{137}$Cs source at the center of the FoV. The exposure time is 13$\;$min. Right three panels show images of gamma rays from three different sources ($^{133}$Ba, $^{137}$Cs, and $^{54}$Mn). The exposure time is 22$\;$min. From left to right, we selected gamma rays around 356, 662, and 835$\;$keV. In the panels, black points show the source positions and source intensity, distance between source and origin, and contribution of sources to dose rate at the center of the drift top of the TPC are shown.}
  \label{fig:gamma-image2}
 \end{center}
\end{figure}

\subsection{Weak-source measurement}
As shown in the left panel of figure \ref{fig:gamma-image2}, we measured 662$\;$keV gamma rays from the $^{137}$Cs source, whose contribution to the dose rate at the center of the drift top of the TPC is 0.15$\;\mu\rm{Sv/h}$, in the $<\;$0.1$\;\mu\rm{Sv/h}$ background dose rate condition for 13 min. If we select gamma rays from inside the yellow circle (polar angle = 20$^{\circ}$) drawn in the figure, we know that the ETCC detected the source image with the significance of 5$\sigma$.

The right three panels of figure \ref{fig:gamma-image2} show images of weak gamma rays from three point sources ($^{133}$Ba (0.52$\;$MBq, 2.07$\;$m), $^{137}$Cs (0.71$\;$MBq, 2.13$\;$m), and $^{54}$Mn (0.43$\;$MBq, 2.03$\;$m)) with several energy selections. The contributions of $^{133}$Ba, $^{137}$Cs, and $^{54}$Mn to the dose rate at the center of the drift top of the TPC are 0.013, 0.017 and 0.016$\;\mu$Sv/h, respectively. The measurement time was 22$\;$min. These figures show clear images of weak sources and were acquired over short times. Because the ETCC can detect gamma rays in sub-MeV and MeV energy bands, we can use it for monitoring environmental gamma rays from $^{134}$Cs and $^{137}$Cs whose primary energies are 605, 662 and 796$\;$keV.

\section{Summary and discussion}
\label{sect:summary}
To measure environmental gamma rays from outdoor soil contaminated with radioactivity, we developed a portable, battery-powered prototype ETCC system with 9 PSAs at the bottom of a $10\;\times\;10\;\times\;16\;$cm$^{3}$ TPC. This system uses a compact readout circuit and the new data-acquisition system developed for SMILE-II FM ETCC. Furthermore, using several laboratory gamma-ray point sources, we verified its capability to image gamma rays and its performance. We obtained clear images of a $^{137}$Cs point source for polar angles from 0$^{\circ}$ to 50$^{\circ}$. We also obtained images of weak gamma rays from the three point sources such as $^{133}$Ba, $^{137}$Cs, and $^{54}$Mn and obtained 356, 662, and 835$\;$keV gamma-ray images that clearly show the positions of the sources on changing the energy-selection range. We checked gamma-ray detection efficiency and ARM as a function of energy and polar angle by performing gamma-ray measurements with several gamma-ray point sources and by performing a simulation. We find that the detection efficiency and the angular resolution for 662$\;$keV gamma rays from the center of the FoV is $(9.31 \pm 0.95) \times 10^{^-5}$ and $5.9^{\circ}\pm0.6^{\circ}$, respectively. From the measurements, we determine that the FoV of the detector is approximately 1$\;$sr. We also verified that the ETCC can detect a 0.15$\;\mu\rm{Sv/h}$ from a $^{137}$Cs gamma-ray source with a significance of 5$\sigma$ within 13 min in the field where the background dose rate is less than 0.1$\;\mu$Sv/h.

Based on these results, we can roughly evaluate the ability of the ETCC to perform environmental gamma-ray measurements for the purpose described in section 1. We consider measuring the area that is uniformly contaminated with $^{137}$Cs at a dose rate of 1$\;\mu$Sv/h. If we divide a gamma-ray image into 0.1$\;$sr bins, which is greater than the angular resolution of the ETCC, we can divide the FoV into ten regions. In this case, the ETCC can detect over ten gamma rays and can measure the gamma-ray rate with approximately 30$\%$ statistical error in each region. This is practical for rapidly checking for hot spot in the contaminated soil. If we use higher-pressure gas or another gas species such as CF$_{4}$ gas in the TPC, the detection efficiency would become several times higher and the measurement time would become even shorter.

\acknowledgments
This research was supported by the Japan Science and Technology Agency, JST. The ETCC technology is based on a technology developed for the SMILE-II experiment, which was supported by a Grant-in-Aid for Scientific Research from the Japan Ministry of Education, Culture, Science, Sports and Technology. We used some electronics equipment that were technically supported by Open-It Consortium.

\end{document}